\documentclass[review]{elsarticle}

\usepackage{lineno,hyperref}
\modulolinenumbers[5]
\usepackage{color}
\usepackage{ulem}
\usepackage{tabularx,ragged2e,booktabs,caption,graphicx,epstopdf}
\usepackage{float}
\journal{Carbon}
\begin{document}
\begin{frontmatter}

\title{Growth of superconducting boron doped diamond on 4inch silicon wafers}

\author[1]{Soumen Mandal\corref{a}}
\ead{mandals2@cardiff.ac.uk}
\author[1]{Oliver A. Williams\corref{a}}
\ead{williamso@cardiff.ac.uk}
\address[1]{Department of Physics and Astronomy, Cardiff University, Cardiff, UK}
\cortext[a]{Corresponding authors}

\begin{abstract}
Superconducting boron-doped diamond (BDD) films were grown on 4-inch silicon wafers by microwave plasma chemical vapour deposition using gas-phase B/C ratios ranging from 6536 to 36421 ppm. Surface morphology, boron incorporation and superconducting properties were investigated as a function of gas-phase boron concentration. No systematic variation in apparent lateral grain size was observed across the series. Superconductivity was observed in all films except that grown at a B/C ratio of 6536 ppm within the measured temperature range down to 2 K. The superconducting transition temperature initially increased with increasing B/C ratio, reaching a maximum T$_c$ of 4.03 K at 24691 ppm, before decreasing at higher gas-phase B/C ratios. The corresponding resistive upper critical field at 2 K reached 3.091 T. Raman spectroscopy showed an increase in boron incorporation with increasing gas-phase B/C ratio up to 30303 ppm, followed by a slight decrease at 36421 ppm. Spatial measurements across the film grown at 24691 ppm showed T$_c$ values of 4.02, 4.19 and 3.33 K at the centre, intermediate and edge positions, respectively, with Raman spectroscopy showing a corresponding spatial variation in boron concentration. Comparison with previous growth on 2-inch wafers showed that substantially higher gas-phase B/C ratios were required to obtain comparable boron concentrations and superconducting properties on 4-inch wafers, indicating reduced boron incorporation efficiency during large-area growth. These results demonstrate the feasibility of producing superconducting BDD over a substantial area of a 4-inch silicon wafer while identifying boron incorporation and radial uniformity as key parameters for further wafer-scale optimisation.
\end{abstract}

\begin{keyword}
boron-doped diamond, superconductivity, chemical vapour deposition, Raman spectroscopy, 4-inch wafer
\end{keyword}

\end{frontmatter}

%\linenumbers

\section{Introduction}
Diamond is an intrinsically insulating allotrope of carbon. However, the incorporation of boron into the diamond lattice introduces acceptor states and increases the hole concentration. With increasing boron concentration, diamond undergoes a transition from semiconducting \cite{sato00} to metallic behaviour \cite{bust08} and, above a critical doping level, becomes superconducting \cite{ekim04} at low temperatures. 

Superconductivity in diamond was first reported by Ekimov et al. in 2004 in heavily boron-doped material synthesised using a high-pressure, high-temperature method \cite{ekim04}. This discovery stimulated considerable interest in diamond and related superhard superconductors.\cite{blas04, bust04, taka05, gaje09, taka09, mand10, dahl10, mand11, baut14, klem21, cuen23} Boron-doped diamond is now established as a type-II superconductor, with the original material exhibiting a critical temperature of approximately 4 K and an upper critical field exceeding 4 T \cite{ekim04}. Subsequent advances in boron incorporation and chemical vapour deposition (CVD) growth have produced homoepitaxial diamond films with superconducting transition temperatures as high as approximately 11 K \cite{taka09}.

Although the nature and microscopic origin of superconductivity in diamond remain under investigation, several devices based on nanocrystalline boron-doped diamond have already been demonstrated, including superconducting quantum interference devices (SQUIDs), superconducting microelectromechanical systems (MEMS) and microwave resonators \cite{mand11,baut14,cuen23}. Most of these devices were fabricated from material grown on 2-inch silicon wafers. Owing to the spatial non-uniformity inherent to the CVD growth process, however, the properties near the wafer edge can differ considerably from those within the central region, thereby limiting the area suitable for device fabrication. The growth of superconducting boron-doped diamond over larger substrates is therefore important for scalable device production. In this study, boron-doped diamond films with varying gas-phase boron-to-carbon ratios were grown on 4-inch silicon wafers. The film morphology was examined using scanning electron microscopy, boron incorporation was evaluated by Raman spectroscopy, and the superconducting properties were investigated through low-temperature electrical-transport measurements.

\section{Materials and Methods}

Boron-doped diamond films were deposited in a Seki Technotron AX6500 microwave plasma-assisted chemical vapour deposition system. Before deposition, the substrates were seeded by brief immersion in an aqueous monodisperse suspension of hydrogen-terminated diamond nanoparticles. The preparation of this seeding suspension has been described previously \cite{hees11}. Film growth was performed at a chamber pressure of 50 Torr and an applied microwave power of 5 kW. The substrate temperature was maintained at approximately 800 $^o${C}, as determined using a Williamson dual-wavelength pyrometer. Methane constituted 3\% of the process-gas mixture, with hydrogen forming the balance. Boron was introduced using a commercially premixed source containing 2000 ppm trimethylboron (TMB) in hydrogen. A series of six films was prepared using gas-phase B/C ratios of 6536, 12,821, 18,868, 24,691, 30,303 and 36,421 ppm, denoted hereafter as 6k, 12k, 18k, 24k, 30k and 36k, respectively. Based on the atomic density of diamond and assuming unity boron-incorporation efficiency, these gas-phase ratios would correspond to nominal boron concentrations of (1.15, 2.26, 3.32, 4.35, 5.33 and 6.41) $\times 10^{21}$ B cm$^{-3}$, respectively. These values represent nominal concentrations derived from the supplied gas composition rather than direct measurements of the boron incorporated into the films.

The surface morphology of the films was examined using a Hitachi SU8200 field-emission scanning electron microscope operated at an accelerating voltage of 20 kV, with working distances ranging from 9 to 11 mm. The apparent lateral grain sizes were estimated from the SEM images using AI-based image-analysis tools.  The films were characterised using a Horiba LabRAM HR Evolution Raman spectrometer equipped with a SynapsePlus back-illuminated deep-depletion CCD detector and a 532 nm excitation laser.  The superconducting properties of the films were investigated using a Quantum Design Physical Property Measurement System (PPMS). Electrical contacts were prepared using silver paste and gold wires in a van der Pauw configuration. Temperature-dependent resistance measurements were performed between 2 and 10 K during warming at a rate of 0.5 K/min to ensure thermal stability. The field-dependent resistance was also measured at 2 K, below the superconducting transition temperatures of the films, to determine their critical fields.

\section{Results and Discussion}
\begin{figure}[!h]
\centering
\includegraphics[width = 9cm]{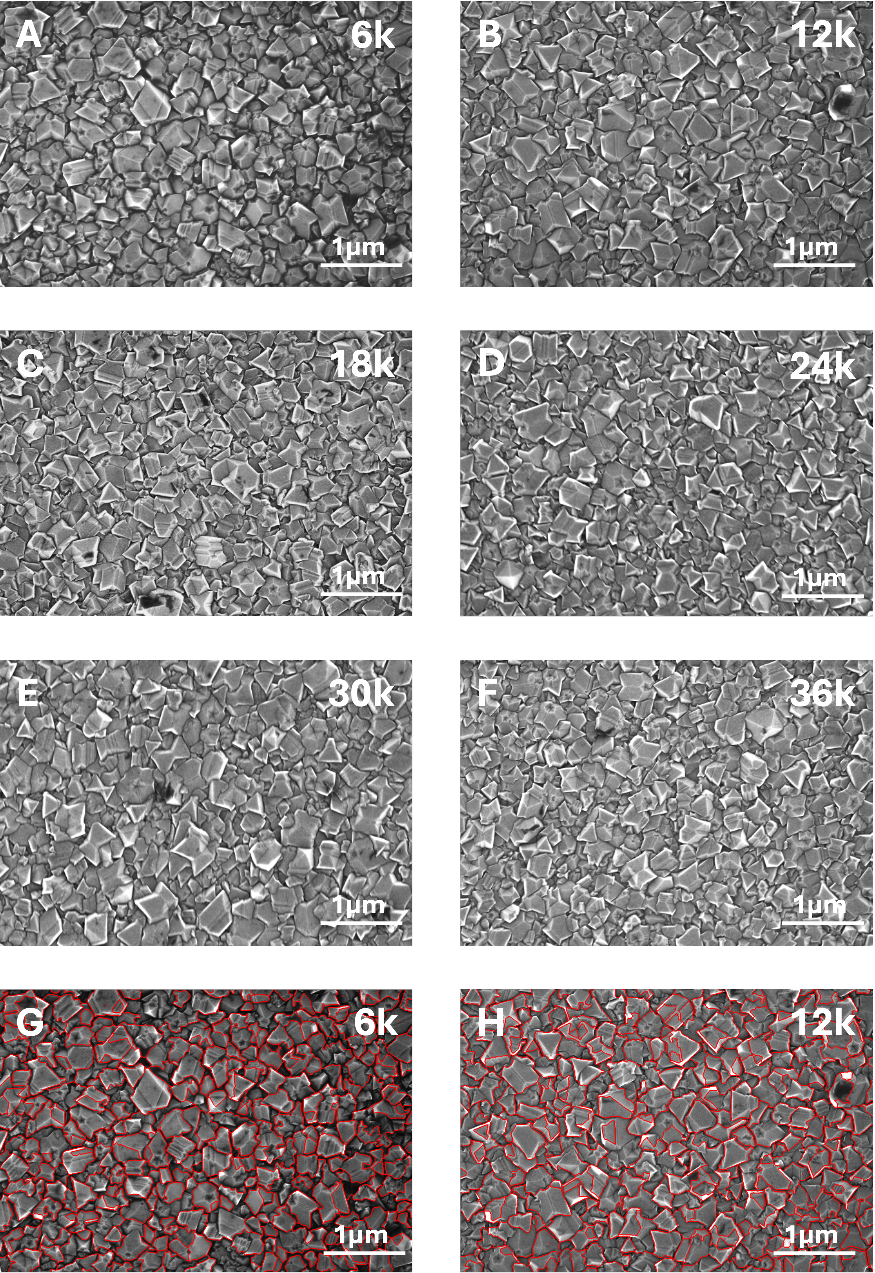}
\caption{SEM images of boron-doped diamond films grown on 4-inch silicon wafers. Panels (a–f) show the surface morphology of films deposited using progressively increasing gas-phase B/C ratios. The sample designation in the upper-right corner of each panel corresponds to the abbreviated gas-phase B/C ratio defined in the main text. No discernible variation in surface morphology is observed across the series. The exact B/C ratio in the gas phase is given in the text. Panels (g) and (h) show representative grain-boundary demarcations for the 6k and 12k samples, respectively. These images illustrate the limitations of automated grain-boundary identification arising from the faceted surface morphology of polycrystalline BDD.} \label{fig1}
\end{figure}

Figures \ref{fig1}(a–f) show representative SEM images of the six boron-doped diamond films, with the sample designation indicated in the upper-right corner of each panel. Visual comparison reveals no substantial change in surface morphology or apparent grain size across the range of gas-phase B/C ratios investigated. The mean apparent lateral grain size of each film was estimated using an automated AI-assisted image-analysis tool. The resulting mean values ranged from approximately 240 to 270 nm, with no systematic dependence on the gas-phase B/C ratio.

It should be noted that the quantitative determination of grain size from SEM images of polycrystalline diamond is inherently challenging. Individual grains frequently contain multiple crystallographic facets and surface growth features, some of which may be incorrectly identified as grain boundaries during automated image analysis. This can divide a single grain into multiple regions and consequently lead to a slight underestimation of the absolute grain size. Nevertheless, all images were acquired at the same magnification and analysed using the same procedure, allowing a meaningful comparison between the samples. Any substantial variation in grain size would also be expected to be apparent from direct visual comparison of the SEM images; no such variation was observed here. Figures \ref{fig1}(g) and (h) show representative boundary demarcations for the 6k and 12k films, respectively. These overlays illustrate the overall reliability of the analysis while also showing the limited misclassification arising from the faceted surface morphology. Therefore, although the calculated values should be regarded as estimates of the apparent lateral grain size, the analysis supports the conclusion that increasing the gas-phase B/C ratio did not produce a systematic change in grain size.

X-ray diffraction can also be used to estimate crystallite size; however, the values obtained represent the dimensions of coherently diffracting domains rather than those of the morphological grains observed by SEM. A single faceted grain may contain twins, defects or sub-grain boundaries and may therefore comprise several smaller coherent domains. Furthermore, diamond growth on silicon begins with nanodiamond seeds, typically 5–10 nm in size \cite{mand21}, which grow and coalesce during deposition to form the larger surface grains visible in Figure \ref{fig1}. Consequently, defining a single mean grain size for polycrystalline BDD is not straightforward. In the present study, the discussion is restricted to the apparent lateral grain sizes determined from plan-view SEM images. No systematic change in these values was observed with increasing gas-phase B/C ratio, indicating that the evolution of the superconducting properties cannot be readily attributed to changes in the surface grain size.

\begin{figure}[!h]
\centering
\includegraphics[width = 7cm]{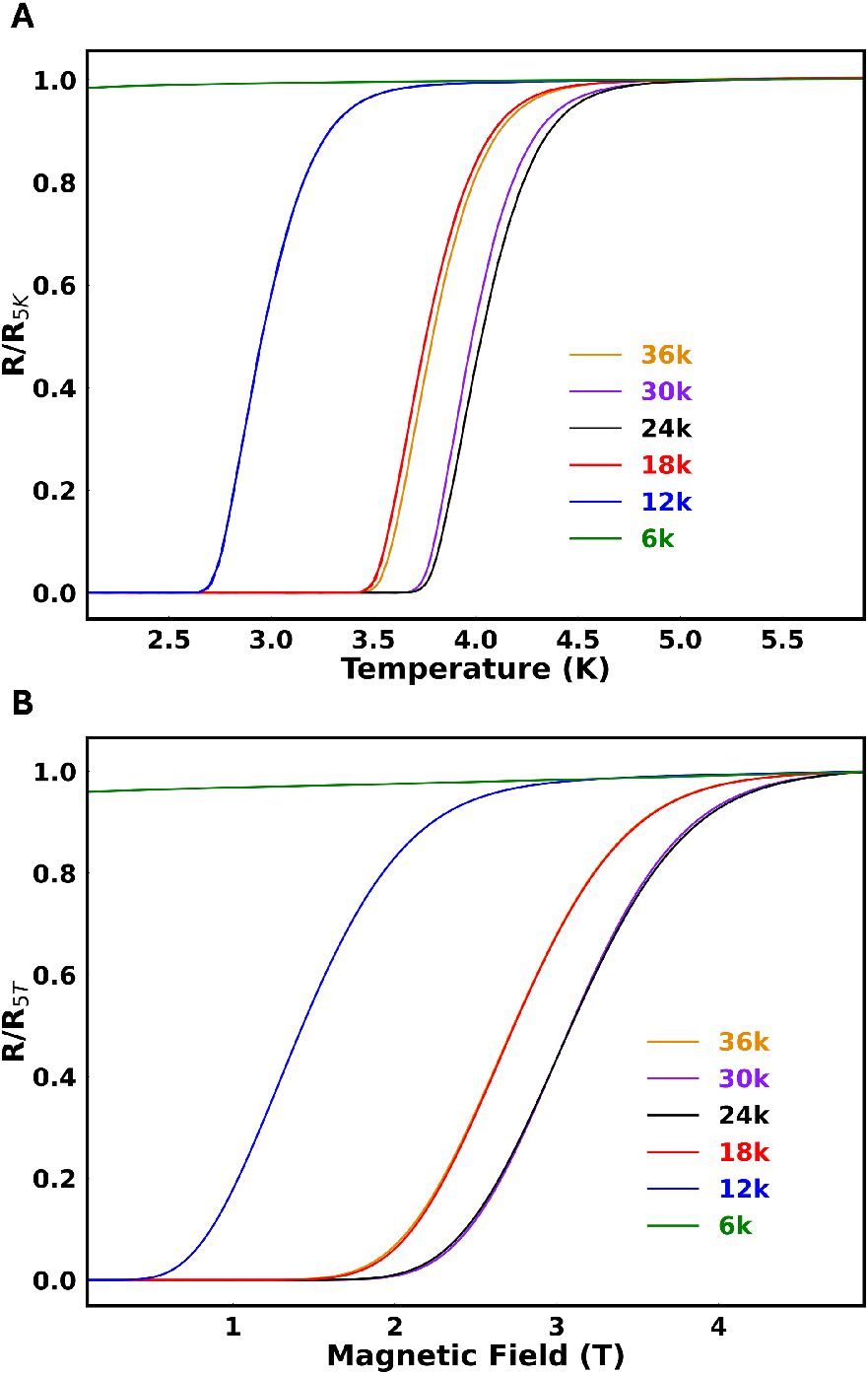}
\caption{(a) Normalised resistance as a function of temperature for the boron-doped diamond films, measured using the van der Pauw method. The resistance was normalised to its value at 5 K. All films except the 6k sample exhibit a superconducting transition above 2 K, with the 24k sample showing the highest T$_c$. (b) Normalised resistance as a function of magnetic field measured at 2 K, with the resistance normalised to its value at 5 T.} \label{fig2}
\end{figure}

Figure \ref{fig2} summarises the superconducting properties of the boron-doped diamond films. Panel (a) shows the temperature dependence of the normalised resistance measured in the van der Pauw configuration. To enable direct comparison between samples, the resistance was normalised to the normal-state resistance at 5 K, R$_n$=R(5 K). The superconducting transition temperature, T$_c$, was defined as the temperature at which the resistance reached 0.5R$_n$. The transition width, $\Delta T_c$, was determined using the 10--90\% criterion and calculated as $\Delta T_c$=T$_{90}$-T$_{10}$, where T$_{90}$ and T$_{10}$ correspond to resistances of 0.9R$_n$ and 0.1R$_n$, respectively. The 6k sample did not exhibit a superconducting transition down to 2 K, while the transition temperature of the 12k sample was substantially lower than those of the more heavily doped films. The highest T$_c$ was observed for the 24k sample.

Panel (b) shows the magnetic-field dependence of the normalised resistance for the same films, measured at 2 K and normalised to the resistance at 5 T. The resistive upper critical fields follow a trend broadly similar to that observed for T$_c$. However, the H$_{c2}$ values of the 24k and 30k samples are nearly identical, as are those of the 18k and 36k samples. By comparison, the temperature-dependent measurements reveal small but measurable differences within these pairs, with the 24k sample exhibiting the highest T$_c$. The 6k sample did not exhibit a superconducting transition down to the minimum measurement temperature of 2 K, while the T$_c$ of the 12k sample was substantially lower than those of the more heavily doped films. In contrast, superconductivity has previously been observed at the centre of BDD films grown on 2-inch silicon wafers using a gas-phase B/C ratio comparable to the 6k condition \cite{benn24}. Moreover, in previous studies on 2-inch wafers, the highest T$_c$ was obtained at a gas-phase B/C ratio corresponding to the 12k condition, with further increases in B/C resulting in a suppression of T$_c$ \cite{mani21}. In contrast, for the present 4-inch films, T$_c$ continues to increase up to the 24k condition, and suppression is observed only at higher gas-phase B/C ratios. These results indicate that the optimum gas-phase B/C ratio for maximising T$_c$ shifts towards substantially higher values when the substrate diameter is increased from 2 to 4 inches. The superconducting transition temperatures, T$_c$, resistively determined upper critical fields, H$_{c2}$, and their corresponding transition widths, $\Delta T_c$ and $\Delta H$, are summarised in Table \ref{tab1}. All values were extracted using the criteria described above.

\begin{table}[htbp]
\centering
\begin{tabular}{ccccc}
\hline
B/C sample &
$T_c^{50\%}$ (K) &
$\Delta T_c$ (K) &
$\mu_0H_{c2}^{50\%}(2 K)$ (T) &
$\Delta H$ (T) \\
\hline
6k  & --   & --   & --    & --    \\
12k & 2.90 & 0.52 & 1.432 & 1.368 \\
18k & 3.75 & 0.52 & 2.738 & 1.411 \\
24k & 4.03 & 0.53 & 3.091 & 1.447 \\
30k & 3.98 & 0.50 & 3.087 & 1.400 \\
36k & 3.79 & 0.52 & 2.734 & 1.423 \\
\hline
\end{tabular}
\caption{Superconducting transition temperatures and resistive upper
critical fields of the boron-doped diamond films. The transition
temperatures were determined from $R(T)/R(5K)$, whereas the
field values were determined at 2~K from $R(H)/R(5T)$.
The transition widths are defined as
$\Delta T_c=T_{90\%}-T_{10\%}$ and
$\Delta H=H_{90\%}-H_{10\%}$.}
\label{tab1}
\end{table}

\begin{figure}[htbp]
\centering
\includegraphics[width = 7cm]{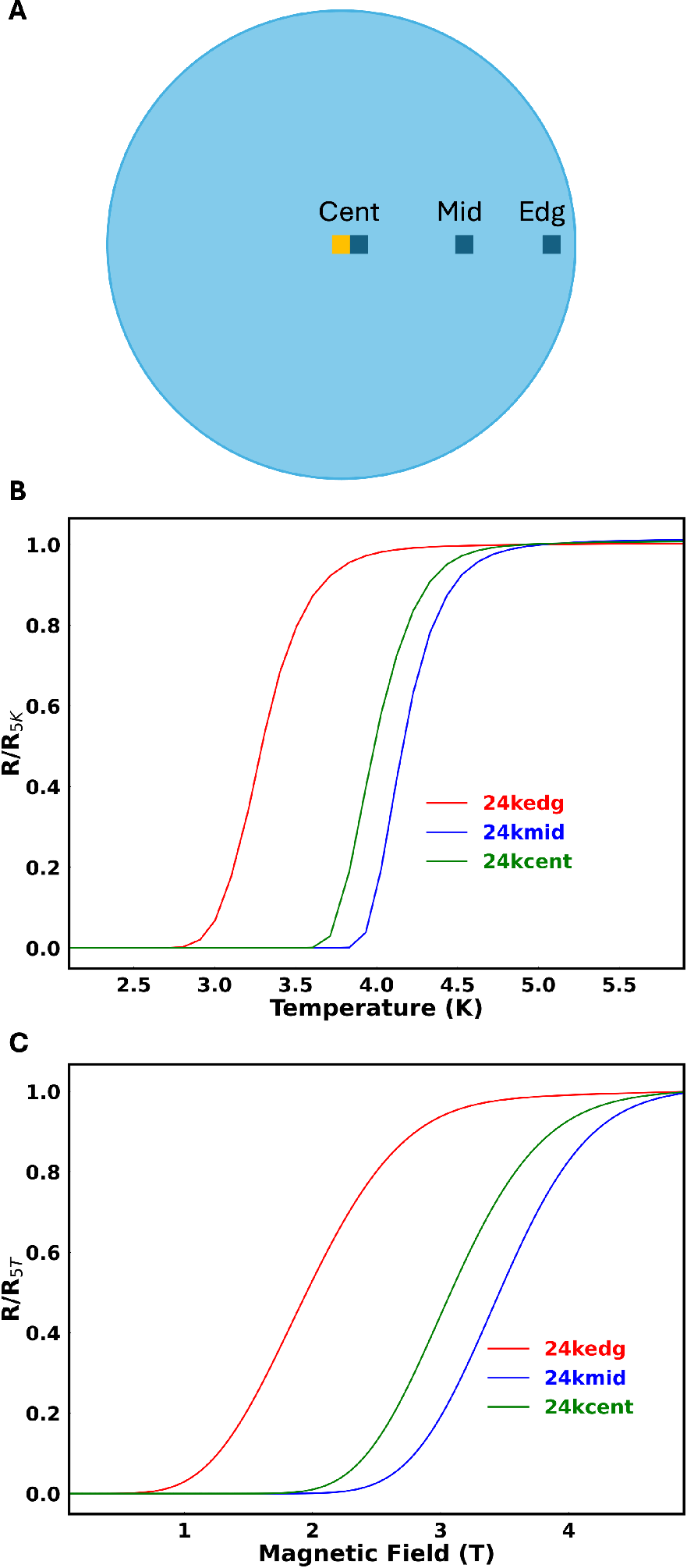}
\caption{(a) Schematic showing the approximate positions of the pieces used for the wafer-uniformity study. The orange region represents the original central piece, for which the superconducting data are presented in Figure \ref{fig2}. (b) Normalised resistance as a function of temperature for pieces taken from different positions on the 24k sample, measured using the van der Pauw method. The resistance was normalised to its value at 5 K. The labels 24kcent, 24kmid and 24kedg correspond to the sampling positions indicated in panel (a). (c) Normalised resistance as a function of magnetic field measured at 2 K, with the resistance normalised to its value at 5 T.} \label{fig3}
\end{figure}

The results presented in Figure \ref{fig2} were obtained from pieces taken from the central region of each 4-inch wafer. To evaluate the area over which the superconducting properties were maintained, the 24k sample, which exhibited the highest $T_c$, was selected for a radial uniformity study. Three additional pieces were examined: a second piece from the central region to assess centre-to-centre reproducibility, and two pieces located 2.3 cm ($\approx 1$ inch) and 4.2 cm from the wafer centre. The latter two pieces were taken along the same radial direction to investigate the evolution of the superconducting properties from the centre towards the wafer edge. These three pieces are hereafter referred to as 24kcent, 24kmid and 24kedg, corresponding to radial distances of 0, 2.3 and 4.2 cm from the wafer centre, respectively. The approximate sample positions are illustrated schematically in Figure \ref{fig3}. Small deviations may exist between the indicated and actual positions on the wafer. Furthermore, because the wafer was diced manually, the resulting pieces were not perfectly square.

Figures \ref{fig3}(b) and (c) show the superconducting properties measured at three positions on the 24k sample. The T$_c$ of 24kcent was 4.02 K, in close agreement with that measured for the original central piece. A modest increase in T$_c$, to 4.19 K, was observed for 24kmid, located 2.3 cm from the centre. Further towards the wafer edge, T$_c$ decreased considerably to 3.33 K for 24kedg. This radial dependence differs somewhat from that previously observed for 2-inch wafers \cite{benn24}, where T$_c$ decreased away from the wafer centre but showed little variation across the central 1-inch region. For the present 4-inch sample, the measurements suggest that a high T$_c$ is maintained across the central region spanning approximately 2 inches in diameter, while the material near the wafer edge remains superconducting but exhibits a lower T$_c$. The resistively determined upper critical fields of these pieces were also measured at 2 K, and the results are shown in Figure \ref{fig3}c. The observed trend closely follows that of the T$_c$ measurements. The central piece exhibits a critical field of 3.079 T, consistent with the value obtained from Figure \ref{fig2} (see Table \ref{tab1}). A slightly higher critical field of 3.459 T is observed for the 24kmid piece, whereas the critical field decreases considerably to 1.954 T at the wafer edge. Taken together, these results demonstrate that superconducting BDD can be grown over large-area wafers, although the gas-phase boron concentration needs to be optimised to achieve the best superconducting properties across the wafer.

\begin{figure}[htbp]
\centering
\includegraphics[width = 6cm]{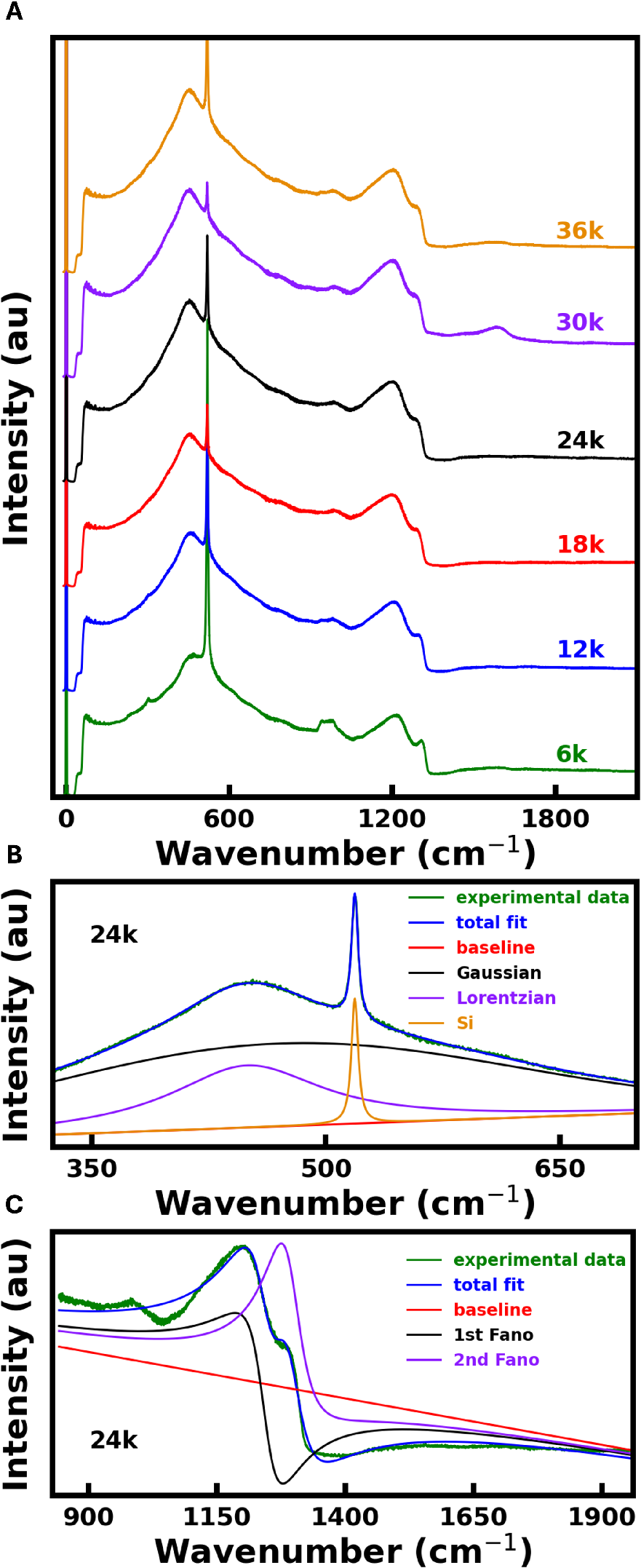}
\caption{(a) Raman spectra of BDD films grown using increasing gas-phase B/C ratios from 6k to 36k. The spectra are vertically offset for clarity. (b) Representative fitting of the low-wavenumber Raman spectrum of the 24k sample. The broad boron-related feature was decomposed into Gaussian and Lorentzian components, while the narrow contribution at approximately 520 cm$^{-1}$, arising from the Si substrate, was fitted separately. The Lorentzian peak position was used to estimate the boron concentration according to the empirical relationship reported by Bernard et al. \cite{bern04}. (c) Double-Fano fitting of the Raman spectrum of the 24k sample using the model proposed by Mortet et al. \cite{mort1, mort2}, showing the experimental spectrum, total fit, fitted baseline and the two Fano components.} \label{fig4}
\end{figure}

While the superconducting properties provide an indication of the boron concentration in the films, Raman spectroscopy was used to independently estimate the boron concentration across the sample series. The Raman spectra of all the samples are presented in Figure \ref{fig4}a. The spectra show the typical characteristics of heavily boron-doped diamond, including a broad feature around 500 cm$^{-1}$ and Fano-like features in the vicinity of 1200--1300 cm$^{-1}$. With increasing gas-phase boron concentration, the diamond zone-centre phonon (ZCP) near 1300 cm$^{-1}$ becomes progressively broader and less well defined, consistent with increased boron incorporation into the diamond films. To estimate the boron concentration, both spectral regions were analysed independently. The low-wavenumber feature was analysed using the method described by Bernard et al.\cite{bern04}, while the higher-wavenumber region was analysed using the online Raman fitting tool based on the double-Fano model developed by Mortet et al.\cite{mort1,mort2,mort3}. A representative double-Fano fit for the 24k sample is shown in Figure \ref{fig4}c. The boron concentrations obtained directly from the online fitting tool were used for comparison with those determined from the low-wavenumber analysis.

For the low-wavenumber region, the broad feature around 500 cm$^{-1}$ was fitted using Gaussian and Lorentzian components after accounting for the Si substrate peak and the spectral background. Bernard et al.\cite{bern04} observed that the position of the Gaussian component showed no systematic relationship with the boron concentration, whereas the position of the Lorentzian component showed a systematic dependence on the boron concentration. The boron concentration was therefore estimated from the fitted Lorentzian peak position using the empirical relationship

\begin{equation}
[B] (cm^{-3})
=
8.44\times10^{30}
\exp\left[
-0.048
\left(
\frac{\omega_L}{\mathrm{cm}^{-1}}
\right)
\right],
\label{eqbor}
\end{equation}

where $\omega_L$ is the position of the Lorentzian component of the low-wavenumber Raman feature. A representative fit of the low-wavenumber Raman data is shown in Figure \ref{fig4}b, using the 24k sample as an example. The same low-wavenumber region was also analysed using the approach reported by Mortet et al. \cite{mort3}, in which the $\sim$500 cm$^{-1}$ feature is described using a Fano lineshape and the boron concentration is related to the position of the fitted feature according to

\begin{equation}
[B]\,(\mathrm{cm}^{-3})
=
-6.99\times10^{19}
\left(
\frac{\omega_{500}}{\mathrm{cm}^{-1}}
\right)
+
3.57\times10^{22},
\label{eqbor2}
\end{equation}

The boron concentrations determined using the three Raman-based approaches, together with the fitted Lorentzian peak positions, are summarised in Table \ref{tab:raman_boron}. For the 6k sample, the boron-related feature around 500 cm$^{-1}$ was too weak relative to the Si substrate contribution to obtain a reliable fit using either of the low-wavenumber methods. Consequently, no boron concentration is reported for this sample using the Bernard or Mortet 500 cm$^{-1}$ methods. The weakness of this boron-related Raman feature is itself consistent with a very low level of boron incorporation in the 6k film.

\begin{table}[htbp]
\centering
\begin{tabular}{cccccc}
\hline
Sample & Gas-phase B/C & Lorentzian centre & \multicolumn{3}{c}{B concentration (cm$^{-3}$)} \\
       & (ppm) & (cm$^{-1}$) & Bernard et al. & Mortet 500 cm$^{-1}$ & Mortet ZCP \\
\hline
6k  & 6536  & --     & --                    & --                    & $4.9\times10^{21}$ \\
12k & 12821 & 455.81 & $2.66\times10^{21}$   & $3.1\times10^{21}$    & $6.1\times10^{21}$ \\
18k & 18868 & 450.96 & $3.35\times10^{21}$   & $3.9\times10^{21}$    & $5.1\times10^{21}$ \\
24k & 24691 & 449.69 & $3.56\times10^{21}$   & $3.9\times10^{21}$    & $7.4\times10^{21}$ \\
30k & 30303 & 448.41 & $3.79\times10^{21}$   & $4.2\times10^{21}$    & $3.9\times10^{21}$ \\
36k & 36421 & 450.31 & $3.46\times10^{21}$   & $3.6\times10^{21}$    & $6.3\times10^{21}$ \\
\hline
\end{tabular}
\caption{Boron concentrations estimated from the Raman spectra using three approaches. The gas-phase B/C ratio used during growth is also shown for comparison. The Bernard et al.\cite{bern04} method uses the position of the Lorentzian component of the broad low-wavenumber feature, while the Mortet 500 cm$^{-1}$ method\cite{mort3} uses the position of the fitted low-wavenumber Fano feature. The Mortet ZCP method\cite{mort1,mort2,mort3} derives the boron concentration from the fitted diamond zone-centre phonon. Reliable low-wavenumber fits could not be obtained for the 6k sample because of the weak boron-related feature and dominant Si substrate contribution.}
\label{tab:raman_boron}
\end{table}

Looking at the superconducting behaviour together with the boron concentrations estimated using the three Raman-based methods, the values obtained using the Bernard et al.\cite{bern04} approach appear to be the most plausible. The Mortet 500 cm$^{-1}$ method shows a broadly similar trend across the sample series, although the absolute concentrations appear to be slightly overestimated when considered alongside the superconductivity data. A larger discrepancy is observed for the concentrations obtained from the Mortet ZCP analysis, particularly for the 6k sample, where a relatively high boron concentration is predicted despite the absence of a superconducting transition down to 2~K. Nevertheless, the absolute concentrations obtained from all three Raman-based approaches should be interpreted with caution, particularly for polycrystalline BDD. A direct measurement of the total boron concentration, for example by secondary ion mass spectrometry (SIMS), was not undertaken in the present study. However, SIMS measures the total boron content and may include boron present at substitutional sites as well as boron incorporated at grain boundaries or other non-substitutional sites. Raman spectroscopy, in contrast, probes boron-induced modifications to the vibrational response of the diamond lattice. The Raman-derived concentrations are therefore used here primarily to assess the relative variation in boron incorporation across the sample series, supported by the corresponding superconducting behaviour, rather than as an absolute determination of the total boron concentration.

\begin{figure}[htbp]
\centering
\includegraphics[width = 6cm]{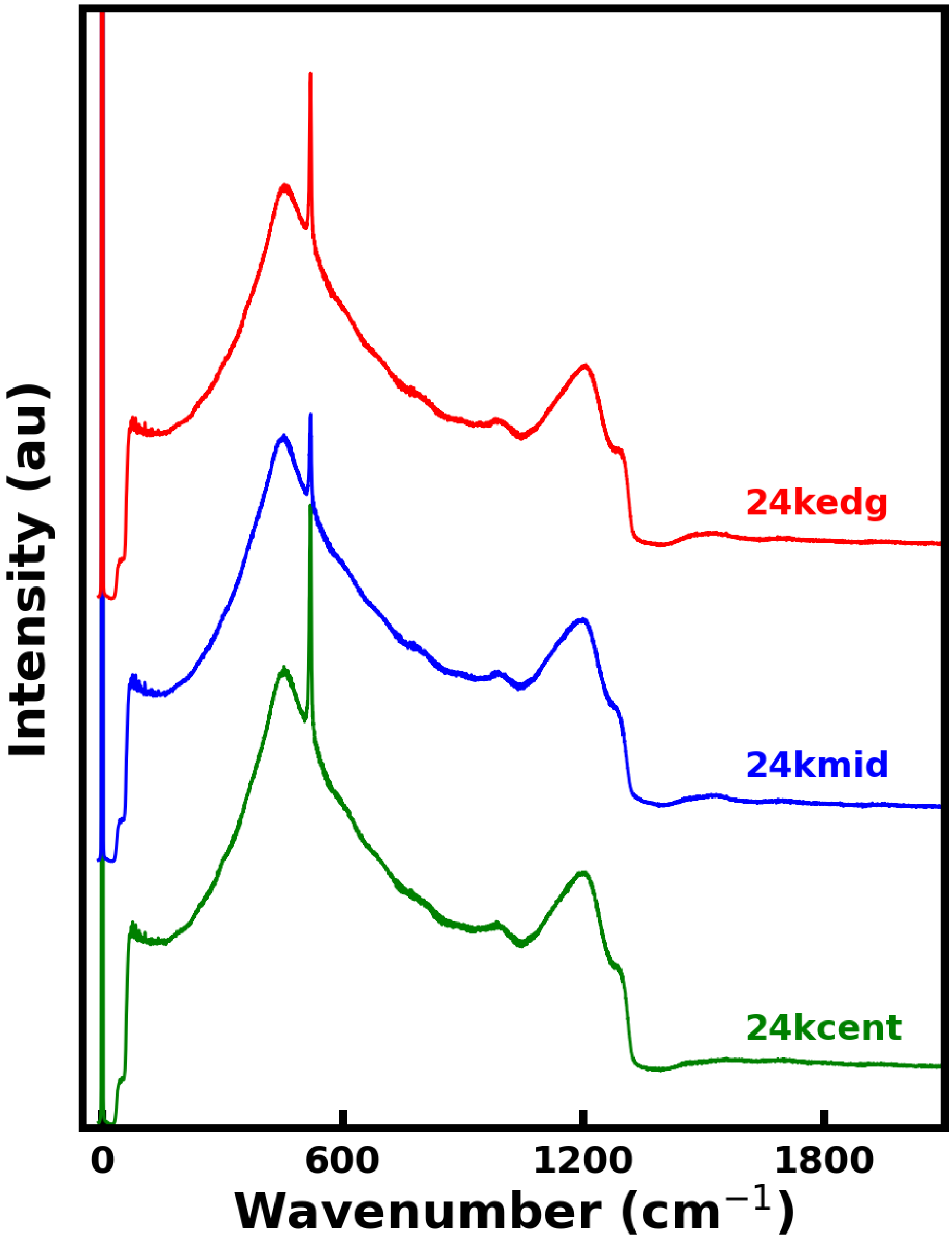}
\caption{Raman spectra of samples 24kcent, 24kmid and 24kedg, taken from the centre, intermediate and edge positions of the 24k BDD wafer, respectively.} \label{fig5}
\end{figure}

Finally, Raman spectroscopy was performed on samples taken from three different positions on the 24k wafer to estimate the spatial variation in boron concentration, with the results shown in Figure \ref{fig5}. Using the $\sim$500~cm$^{-1}$ feature and the Bernard et al.\cite{bern04} method, the boron concentrations were estimated to be $3.56\times10^{21}$, $4.27\times10^{21}$ and $3.01\times10^{21}$~cm$^{-3}$ for the 24kcent, 24kmid and 24kedg samples, respectively. This variation is consistent with the superconducting behaviour, with the 24kmid sample exhibiting the highest T$_c$ and critical field, followed by 24kcent, while both decrease considerably towards the wafer edge.

Comparing the 24kmid and 30k samples, their boron concentrations determined using the Bernard method are relatively similar, although the T$_c$ of the 24kmid sample is marginally higher. Such differences may arise from a number of factors, including variations in the concentration of electrically active substitutional boron, strain, microstructure, and uncertainties associated with estimating the boron concentration from Raman spectroscopy, as discussed above. Hall measurements would provide a more direct determination of the carrier concentration and hence additional information on the electrically active boron. However, Hall measurements of BDD films grown on doped Si substrates can be complicated by parallel conduction through the substrate, potentially resulting in erroneous values. One possible approach would be to grow thin diamond films on electrically insulating 4-inch quartz substrates. This, however, presents a significant growth challenge because of the large thermal-expansion mismatch between diamond and quartz, making the growth of adherent films difficult. Consequently, growth conditions optimised for 4-inch Si substrates may not be directly transferable to 4-inch quartz.

Comparison with the 2-inch wafers studied by Bennett et al.\cite{benn24} suggests that the boron incorporation efficiency during 4-inch growth is considerably lower. For the 24k sample, the boron concentrations derived using the Bernard and Mortet 500~cm$^{-1}$ methods are in relatively good agreement, as shown in Table \ref{tab:raman_boron}. Bennett et al.\cite{benn24} reported boron concentrations of $4.13\times10^{21}$, $3.53\times10^{21}$ and $3.52\times10^{21}$~cm$^{-3}$ for samples with T$_c$ values of 3.86, 3.77 and 3.05~K, respectively. These samples were obtained from a film grown using a gas-phase B/C ratio of approximately 12,800~ppm. Comparable boron concentrations and superconducting properties in the present study are obtained for the 24k and 30k films, which required substantially higher gas-phase B/C ratios of 24,691 and 30,303~ppm, respectively. This comparison therefore supports a reduced boron incorporation efficiency when scaling the growth process to 4-inch wafers. 

The reduced boron incorporation observed during 4-inch growth may be related to the lower effective plasma power density associated with the substantially larger growth area. For the 2-inch wafer growth reported by Bennett et al.\cite{benn24}, a microwave power of 3.5 kW and a pressure of 40 Torr were used, whereas the present 4-inch films were grown at 5 kW and 50 Torr. Although the applied microwave power was increased, the substrate area increases by a factor of four when moving from a 2-inch to a 4-inch wafer. Furthermore, as the plasma occupies a three-dimensional volume, maintaining comparable plasma conditions over the substantially larger growth area would potentially require considerably higher microwave power and/or pressure. Such conditions may not be practically achievable within the constraints imposed by the reactor geometry, available microwave generator power and substrate cooling capacity. Increasing the gas-phase B/C ratio therefore provides a more practical route to achieving the boron incorporation required for higher T$_c$ during 4-inch growth.

 Importantly, the SEM observations indicate that the increased gas-phase B/C ratios used here do not result in an obvious deterioration of the surface morphology of the films. Nevertheless, the present results demonstrate that superconducting BDD can be grown over a substantial area of a 4-inch Si wafer, providing considerably more material for device fabrication and other applications.

\section{Conclusion}
This work demonstrates the growth of superconducting boron-doped diamond over a substantial area of a 4-inch silicon wafer and examines the relationships between gas-phase B/C ratio, film morphology, boron incorporation and superconducting behaviour. Superconductivity was observed in all films except the 6k sample within the measured temperature range down to 2 K. The T$_c$ initially increased with increasing gas-phase B/C ratio, reached a maximum at an intermediate boron concentration, and subsequently decreased for the 30k and 36k samples. This non-monotonic behaviour shows that increasing the boron supply beyond an optimum level does not necessarily enhance the superconducting properties.

Raman spectroscopy was used to estimate the boron incorporation across the sample series using three different approaches. The Bernard et al. and Mortet 500 cm$^{-1}$ methods showed broadly consistent trends, with the estimated boron concentration increasing with gas-phase B/C ratio up to the 30k sample before decreasing slightly for 36k, while the ZCP-based method showed considerably greater variation. Spatial measurements of the 24k film revealed superconductivity at the centre, middle and edge of the wafer, with T$_c$ values of 4.02, 4.19 and 3.33~K, respectively. Raman analysis of the same positions showed a corresponding variation in boron concentration, with the highest concentration observed for 24kmid, followed by 24kcent and 24kedg. Although the lower T$_c$ and boron concentration towards the wafer edge indicate radial non-uniformity, the measurements along the selected radial direction indicate that strong superconducting properties are maintained across the central region spanning approximately 2 inches in diameter..

Compared with previous growth on 2-inch wafers, substantially higher gas-phase B/C ratios were required to achieve comparable Raman-estimated boron concentrations and superconducting behaviour on the 4-inch wafer, suggesting a reduced boron incorporation efficiency during large-area growth. This may be associated with a lower effective plasma power density over the larger growth area, although further investigation is required to establish the underlying mechanism. Overall, the results demonstrate the feasibility of producing superconducting BDD over a substantial area of a 4-inch silicon wafer, providing considerably more material for device fabrication and other applications while identifying the control of boron incorporation and radial uniformity as important challenges for further wafer-scale development.

\section*{Acknowledgment}
SM and OAW would like to acknowledge funding from the UK EPSRC grant "Next generation Acoustic Wave Filter Platform" (No. EP/W036827/1).  The raw data for the results presented here, including how to access them, can be found in the Cardiff University data catalogue at

%\section*{References}
\bibliography{ref}
\end{document}